\documentclass[12pt,final]{article}  

\usepackage{graphicx} 

\input besscleo.sty 

\input besscleo.defs 

\begin{document}

\def\3S1{$^3{\rm S}_1$ }
\def\1S0{$^1{\rm S}_0$ }
\def\L{\rm L}
\def\S{\rm S}
\def\L{\rm L}
\def\J{\rm J}
\def\B{\rm B}
\def\D{\rm D}
\def\P{\rm P}
\def\K{\rm K}
\def\bd{\begin{displaymath}}
\def\ed{\end{displaymath}}
\def\be{\begin{equation}}
\def\ee{\end{equation}}

\title{Charmonium at BES and CLEO-c}
\author{T.Barnes}
\address{Physics Division ORNL, Oak Ridge, TN 37831, USA\\
Department of Physics and Astronomy, University of Tennessee, 
Knoxville TN 37996, USA}
\email{tbarnes@utk.edu}
\maketitle

\abstract{This paper gives a short summary of some of the aspects of
charmonium which can be addressed at BES and CLEO-c and other 
$e^+e^-$ facilities. These topics include
the spectroscopy of charmonium states, radiative transitions, 
$e^+e^-$ widths, 
two-photon widths,
hadron loop effects and open-flavor strong decays.}

\section{Introduction}

Charmonium has been called the ``hydrogen atom of QCD", since many of the
most characteristic and interesting aspects of QCD can be inferred from
studies of the spectrum of charmonium states and their decays and interactions. 
Charmonium is a useful system for the study of forces between quarks in QCD, 
especially the poorly understood, nonperturbative 
confining interaction. This system is of special interest 
because the dynamics are only quasirelativistic, and the 
quark-gluon coupling at this mass scale is intermediate in strength;
for this reason the more unusual features of QCD, such as 
relativistic corrections to the properties of bound states, spin-dependent
forces from quark motion, virtual meson decay loops and various
other effects may be only moderately large ``controlled" 
corrections to a simple nonrelativistic potential model picture. 
Thus in charmonium we have a laboratory in which various novel 
dynamical effects may simultaneously be large enough to be 
clearly identified, and small enough so that they can be treated 
as perturbations of a familiar quantum mechanical model.   

In this contribution we will discuss some of the aspects of charmonium that 
are most easily accessible at BES and CLEO-c. 
We will primarily discuss charmonium states above the open-charm 
threshold of 3.73~GeV, since these are quite poorly known; it 
should be straightforward 
to improve our knowledge in this area 
considerably. For our discussion we use a ``generic" 
nonrelativistic potential model to generate the spectrum of states, 
and the resulting wavefunctions are used to evaluate electromagnetic
transitions involving low-lying charmonium levels.
We then discuss open-flavor strong decays of charmonium states, which 
is an important and largely unexplored topic. 
In our discussion we consider the properties of $1^{--}$ states 
in particular, as these are the most easily accessed at $e^+e^-$ facilities.

\section{A Generic Charmonium Potential Model}

There are several charmonium potential models in the literature, which for the 
most part share the common features of short-ranged color Coulomb
and long-ranged linear scalar confining potentials. In addition to this standard
``Coulomb plus linear" potential, we also include a
Gaussian-smeared spin-spin hyperfine term in our ``zeroth-order potential", 
since a careful treatment of the spin-spin force is important for the
determination of the \3S1 - \1S0 mass splittings. 
This issue has attracted recent attention due to the discovery
of the $\eta_c{\, '}$ by BELLE 
\cite{Choi:2002na}
at a surprisingly high mass 
of $3654\pm 6 \pm 8 $~MeV
(their average including inclusive production is $3639\pm 6$~MeV), 
since confirmed by BABAR  
\cite{Aubert:2003pt}
(at $3630.8\pm 3.4 \pm 1.0 $~MeV)
and CLEO 
\cite{unknown:2003bk}
(at $3642.9\pm 3.1 \pm 1.5 $~MeV).
The zeroth-order potential model Hamiltonian we assume is of the form
\be
{\rm H}_0 = \frac{\vec p^{\; 2}}{m_c}
-\frac{4}{3}\frac{\alpha_s}{r} + br
+
\frac{32\pi\alpha_s}{9 m_c^2}\,
\tilde \delta_{\sigma}(r)\,
\vec {\rm S}_c \cdot \vec {\rm S}_{\bar c}\,
\ee
where 
$\tilde \delta_{\sigma}(r) = (\sigma^3/\pi^{3/2}) \exp{(-\sigma^2 r^2)}$.
Solution of the Schr\"odinger equation with this H$_0$ gives our zeroth-order
charmonium wavefunctions. Splittings within an orbital multiplet are then
determined by taking the matrix element of the spin-dependent Hamiltonian
H$_{\rm I}$ between these zeroth-order wavefunctions. 
The spin-dependent Hamiltonian is taken from the one-gluon-exchange 
Breit-Fermi Hamiltonian (which gives spin-orbit and tensor terms) and
an inverted spin-orbit (Thomas precession) term, which follows from the 
assumption of a Lorentz scalar confining interaction. This H$_{\rm I}$
is explicitly
\be
{\rm H}_{\rm I} = 
\frac{2 \alpha_s}{m_c^2r^3}\, \vec {\rm L}\cdot \vec {\rm S}
+
\frac{4\alpha_s}{m_c^2r^3}\, {\rm T}
-
\frac{b}{2m_c^2r}\, \vec {\rm L}\cdot \vec {\rm S}
\ .
\ee
The assumption of a Lorentz scalar confining interaction is of course 
questionable; the nature of the spin-dependent contributions of the 
confining interaction to the interquark force is one of the interesting open
questions which we hope to address in the study of charmonium. 
 
There are four parameters in this simple potential model,
the one-gluon-exchange coupling constant $\alpha_s$ 
(here for simplicity taken to be a constant), the string tension $b$, the 
charm quark mass $m_c$, and the contact hyperfine interaction smearing 
parameter $\sigma$. We have fixed these using an equal-weight fit to the
the masses of 
the 11 reasonably well established experimental charmonium states; 
the results are
$\alpha_s = 0.5461$, $b = 0.1425$~GeV$^2$, $m_c = 1.4794$~GeV and 
$\sigma = 1.0946$~GeV, which gives an rms error of 13.6~MeV. 
(Of course no special meaning should be attached to 
these precise values, since the model is somewhat {\it ad hoc} and is at best
only an approximate description of the physics of charmonium.)
The experimental states used as input and the predictions for the 
$c\bar c$ spectrum (states below 4.2~GeV and all known $1^{--}$ states)
are given in \Tab{spec_table}; these and some additional levels are
shown in Fig.1.

\Begfigure{t} 
\includegraphics[width=5.5in]{bes1.eps}
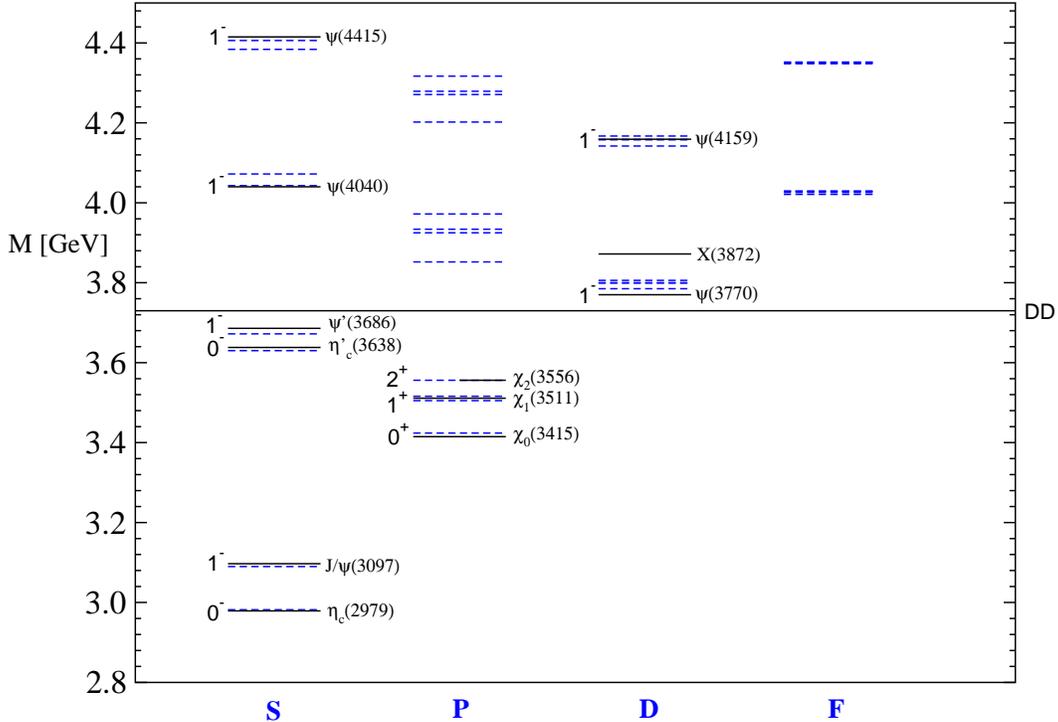
\Endfigure{The spectrum of charmonium states predicted by
the $c\bar c$ potential model described in the text, together with the 
11 known experimental states. The X(3872) is also shown, although its 
identification with $c\bar c$ appears dubious.}{fig_spec}

It is evident in Fig.1 that the simple Coulomb plus linear potential model 
gives a very satisfactory fit to the known charmonium states. It is 
remarkable that both the mean level positions and the splittings within 
the 1S, 1P and 2S multiplets are well described, since they are 
testing different aspects of the interquark forces. The mean multiplet 
positions test the accuracy of the ``funnel-shaped" potential, 
and determine approximate values for the two parameters $\alpha_s$ and $b$. 
The multiplet splittings in contrast 
test whether the spin-dependent interactions 
(spin-spin, spin-orbit and tensor) are also well described with the same 
parameters. These terms arise from the assumptions of transverse 
one-gluon exchange at short distances (which provides the $J/\psi$-$\eta_c$ and 
$\psi{\, '}$-$\eta_c{\, '}$ spin-spin splitting and dominates the P-wave splitting), 
and scalar confinement (which provides a negative spin-orbit term that is 
required for good agreement with the experimental $\chi_{\rm J}$ masses). 
The general good agreement between the potential model and experiment 
is remarkable in view of the additional, presumably large effects such as 
virtual decay loops that have not been included in the model.

We note in passing that lattice QCD has also been applied to charmonium, 
and predicts a very similar spectrum. 
(See for example the recent work of Liao and Manke
\cite{Liao:2002rj}.) 
It is very encouraging that the theoretically better-justified LQCD approach 
gives a pattern of energy levels which is very similar to the potential 
model; this supports our use of the more intuitive potential model to describe 
transitions involving charmonium states in this paper. 
We shall note however in our discussion of decays that virtual decay loops 
are expected to have 
important effects on both the spectrum and the composition 
of charmonium states \cite{vanBeveren:bd,Heikkila:1983wd},
and these effects are neglected in pure $c\bar c$ potential models 
as well as in quenched LQCD.

\section{Electromagnetic couplings}

\subsection{E1 radiative transitions}

Radiative transitions are a very interesting feature of charmonium physics. 
They are quite straightforward to evaluate in $c\bar c$ potential models, and 
(with sufficient statistics) provide a route from the initial 
$1^{--}$ states produced in $e^+e^-$
annihilation to C = $(+)$ charmonia.

The largest rates are for E1 (electric dipole) transitions, which we calculate using

\be
\Gamma_{\rm E1}(
{\rm n}\, {}^{2{\S}+1}{\rm L}_{\J}
\to
{\rm n}'\, {}^{2{\S}'+1}{\rm L}'_{{\J}'}
+ \gamma)
 =  \frac{4}{3}\,  e_c^2 \, \alpha \,
{\rm E}_{\gamma}^3 \,
\frac{{\rm E}_f^{(c\bar c)}}{{\rm M}_i^{(c\bar c)}}\,
C_{fi}\,
\delta_{{\S}{\S}'} \,
|\,\langle
{\rm n}'\, {}^{2{\S}'+1}{\rm L}'_{{\J}'}
|
\; r \;
|\,
{\rm n}\, {}^{2{\S}+1}{\rm L}_{\J}
\rangle\, |^2
\ .
\ee
Here
$e_c= 2/3$ is the $c$-quark charge in units of $|e|$,
$\alpha$ is the fine-structure constant,
E$_{\gamma}$ is the photon energy, and the angular matrix element
$C_{fi}$ is
\be
C_{fi}=\hbox{max}({\L},\; {\L}') (2{\J}' + 1)
\left\{ { {{\L}' \atop {\J}} {{\J}' \atop {\L}} {{\S} \atop 1}  } \right\}^2 .
\ee
This formula is the same as that quoted by 
Ref.\cite{Kwong:1988ae},
except for our inclusion of a relativistic phase space factor of
${\rm E}_f^{(c\bar c)}/{\rm M}_i^{(c\bar c)}$. 
(This is usually not far from unity for the
cases we consider.)
To generate numerical results for transitions of particular interest to
BES and CLEO-c, we have evaluated the matrix elements
$\langle {n'}^{2{\S}'+1}{\L}'_{{\J}'} |\; r \;
| n^{2{\S}+1}{\L}_{\J}  \rangle$
using the nonrelativistic Schr\"odinger wavefunctions of the model
described in the previous section. The resulting E1 rates are given in
\Tab{rad_table}, together with the current experimental numbers.

Several very interesting features of E1 radiative transitions are evident 
in \Tab{rad_table}. First, the 
1P $\to$ 1S
transitions are in very reasonable agreement with experiment.
It is notable that the predicted radiative partial width for 
$h_c \to \gamma \eta_c$ is especially large, which suggests that 
decays to $\gamma \eta_c$ may provide a discovery 
channel for the elusive $h_c$. One possibility is 
$\gamma\gamma \to \eta_c{\, '}$, followed by the decay chain
$\eta_c{\, '} \to \gamma h_c$, $h_c \to \gamma \eta_c$. 
Since these electromagnetic
couplings are all reasonably well understood, detection of the $h_c$ would
simply be a matter of accumulating adequate statistics at a high-energy 
$e^+e^-$ facility with sufficient $\gamma\gamma$ luminosity at 
$\sqrt{s} = 3.7$~GeV.

The theoretical rates for the 
2S $\to$ 1P transitions appear too large by about a factor of two, 
although the relativized model of Godfrey and Isgur \cite{Godfrey:xj}
does not share this difficulty. We note in passing that good agreement
between a pure-$c\bar c$ charmonium potential model and experiment 
may be spurious; decay loop effects will contribute two-meson continuum 
components to all these charmonium resonances, which may significantly 
modify the predicted radiative transition rates.

E1 radiative transitions from the higher-mass charmonium states are
especially interesting. The 1${}^3$D$_1$ candidate $\psi(3770)$ is predicted to
have large partial widths to $\gamma \chi_1$ and  $\gamma \chi_0$
(with branching fractions of 0.5\% and 1.7\% respectively), 
but the branching fraction to $\gamma \chi_2$ is predicted to be 
only about $2 \cdot 10^{-4}$. This small number however follows 
from the assumption that the $\psi(3770)$ is a pure ${}^3$D$_1$ state;
if there is a significant admixture of S-wave basis states in the 
$\psi(3770)$,
\be
|\psi(3770)\rangle =
\cos(\theta) \, |{}^3{\rm D}_1 \rangle
+
\sin(\theta) \, | {2}^3{\rm S}_1 \rangle
\label{3770_mix}
\ee
one typically finds a much larger radiative width to $\gamma \chi_2$
\cite{Rosner:2001nm,Rosner:2004mi}.
(See \Fig{3770_rad}. The sign of the mixing angle
$\theta$ depends on the convention for the normalization of
the ${}^3$D$_1$ and ${2}^3$S$_1$ basis states; note for example
in Fig.1 of Ref.\cite{Rosner:2004mi} that a zero 
$\psi(3770) \to \gamma \chi_2$ width requires 
a small negative mixing angle,
whereas with our conventions it would be positive.)
Since the coupling of the $\psi(3770)$ to $e^+e^-$ suggests a 
significant $2^3$S$_1$ component, a measurement of this radiative
partial width will be especially useful as an independent test of
the presence of this amplitude in the $\psi(3770)$ wavefunction.

We note in passing that if the dominant mechanism of 
${}^3$D$_1$ - ${2}^3$S$_1$ basis state mixing in the 
$\psi(3686)$ and $\psi(3770)$ is through virtual charm meson decay loops
such as DD, DD$^*$ and D$^*$D$^*$, the assumption of a 
$2\otimes 2$ orthogonal mixing matrix as in Eq.\ref{3770_mix} is incorrect.
In this case the
$\langle {}^3$D$_1|\psi(3686)\rangle $ 
and 
$\langle {2}^3$S$_1|\psi(3770)\rangle $
overlaps will no longer be simply related, 
and radiative transition amplitudes will also 
receive contributions from photon emission from the two-meson continua. 

\Begfigure{t} 
\includegraphics[width=5.5in]{3770_rad.eps}
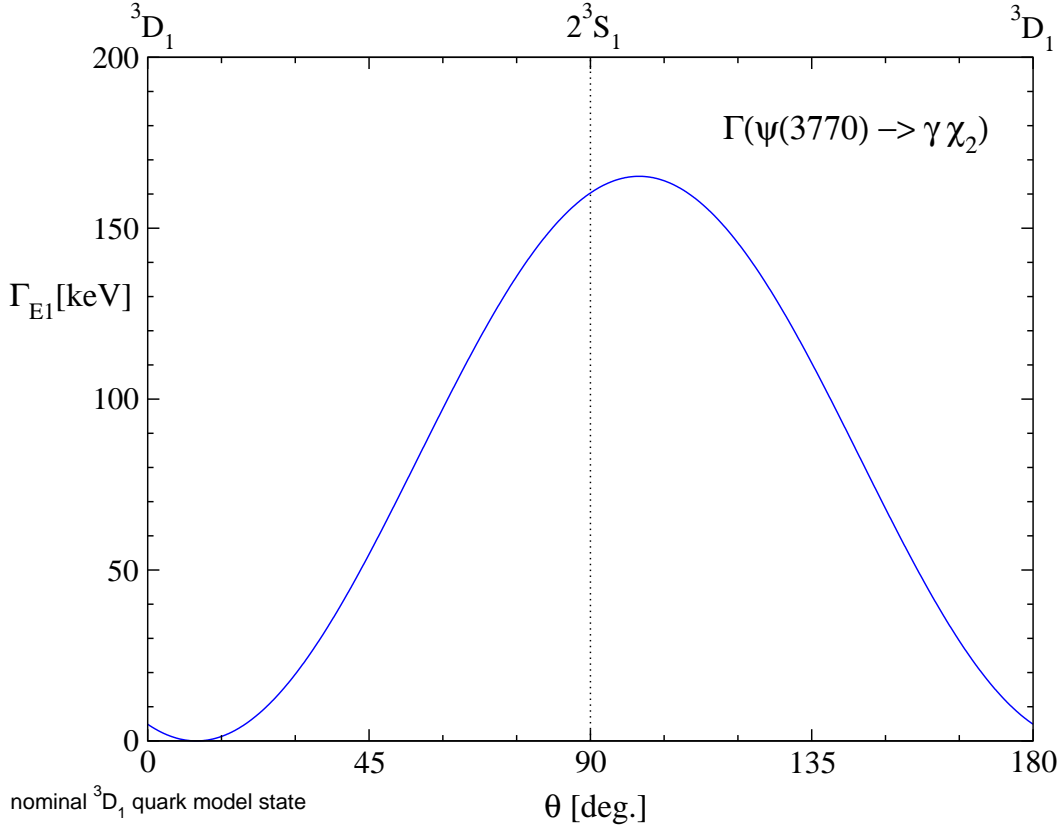
\Endfigure{Predicted radiative partial width for the E1 transition
$\psi(3770)\to \gamma \chi_2$ as a function of the 
${2}^3$S$_1$-${}^3$D$_1$ mixing angle $\theta$.}
{3770_rad}
 
Next we consider radiative decays of the higher-mass vectors
$\psi(4040)$ and $\psi(4159)$.
As is evident in \Tab{rad_table}, if the $\psi(4040)$ is dominantly a 
${3}^3$S$_1$ state as we assume here, it should have very small E1 radiative
widths to the triplet members of the 1P multiplet, with branching fractions 
of at most about $10^{-5}$. The radiative widths to the unknown 2P triplet 
states are theoretically much larger, with branching fractions of  
0.3-1.$ \cdot 10^{-3}$. Since the 2P states have large 
strong branching fractions to DD ($\chi_0$ and $\chi_2$) and
DD$^*$ ($\chi_1$ and $\chi_2$) \cite{Bar04}, 
it may be possible to identify these states
in the DD and DD$^*$ invariant mass distributions of 
$\psi(4040) \to \gamma $DD and $\gamma $DD$^*$ decays.  

Radiative decays of the $\psi(4159)$ share certain features with the 
decays of both the $\psi(3770)$ and the $\psi(4040)$.  
As we found for the $\psi(4040)$, the E1 coupling of the $\psi(4159)$
to the 2P multiplet is much stronger than to the 1P multiplet,
so radiative decays of the $\psi(4159)$ could be used to search for 2P 
states. The branching fraction to the ${2}^3$P$_0$ $\chi_0$ state in particular
is predicted to be a relatively large 0.6\%. 
A strong suppression of $\psi(4159)$ E1 decays to n${}^3$P$_2$ states
is predicted, as we found previously for the $\psi(3770)$. 
This again follows from the assumption that the state is  
purely D-wave $c\bar c$. Given an important S-wave component, 
which is suggested by the $e^+e^-$ width, the coupling 
of the $\psi(4159)$ to n${}^3$P$_2$ states should be much larger. 
Finally, a new feature is that one
may reach the F-wave multiplet (specifically the ${}^3$F$_2$ state)
through the radiative decay of the $\psi(4159)$. Unfortunately this 
${}^3$F$_2$ state is expected to be rather broad; we anticipate a total
width of about 160~MeV, mainly to DD \cite{Bar04}. 

\vskip 0.5cm

\subsection{M1 radiative transitions}

M1 transitions between charmonium states
in pure $c\bar c$ models result from photon emission through the
$H_I = - \vec \mu \cdot \vec B$ magnetic moment interaction
of the $c$ quark (and antiquark), and as such are suppressed relative to
E1 transitions by the small factor of $1/m_c$ in the magnetic
moment operator. The M1 transition amplitude is proportional to the matrix
element of the spin operator, with a spatial factor that
(without recoil corrections) is simply the
matrix element of unity. M1 transitions are therefore nonzero only
between states with the same L$_{c\bar c}$
(and different S$_{c\bar c}$, since the C-parity must change).
If we assumed a spin-independent zeroth-order potential and neglect
recoil effects, M1 transitions between different radial multiplets
would vanish because the
n$^3$S$_1$
and
n$'^1$S$_0$
states have orthogonal spatial wavefunctions. One such transition
is actually observed in charmonium,
$\psi{\, '} \to \gamma \eta_c$,
which must be due in part to the nonorthogonal $\psi{\, '}$ and $\eta_c$
spatial wavefunctions and final meson recoil effects.
                                                                                The formula for M1 decay rates analogous to the E1 formula used 
in the previous section is

\bd
\Gamma_{\rm M1}(
{\rm n}\, {}^{2{\S}+1}{\rm L}_{\J}
\to
{\rm n}'\, {}^{2{\S}'+1}{\rm L}'_{{\J}'}
+ \gamma)
 =  \frac{4}{3}\,  e_c^2 \, \frac{\alpha}{m_c^2}\,
{\rm E}_{\gamma}^3 \,
\frac{{\rm E}_f^{(c\bar c)}}{{\rm M}_i^{(c\bar c)}}\,
{2\J'+1\over 2\L +1}\,
\delta_{{\L}{\L}'} \,
\delta_{{\S},{\S}'\pm 1} \,
\ed
\be
\cdot
|\,\langle
{\rm n}'\, {}^{2{\S}'+1}{\rm L}'_{{\J}'}
|\,
{\rm n}\, {}^{2{\S}+1}{\rm L}_{\J}
\rangle\, |^2
\ .
\ee
Evaluating this formula for transitions from the $\psi$ and
$\psi{\, '}$ gives
the results shown in \Tab{M1rates}. A more detailed study of M1 radiative
decay rates, incorporating recoil corrections (which are numerically 
important for transitions between multiplets such as 
2S $\to$ 1S) will appear in Ref.\cite{Bar04}.
                                                                                    
\Begtable{ht}
{Theoretical and experimental M1 radiative partial widths
of the $\psi$ and $\psi{\, '}$, neglecting recoil effects.}
{M1rates}
\Begtabular{ccccc}
Initial state & Final state
& \phantom{xxxx}  &  $\Gamma_{thy.}$~(keV) & $\Gamma_{expt.}$~(keV) \\
\hline
J$/\psi$   & $\gamma \eta_c $           && 2.9   &  $ 1.2 \pm 0.3$  \\
\hline
$\psi{\, '}$   &  $\gamma \eta_c{\, '}$          && 0.21  &                   \\
$\psi{\, '}$   &  $\gamma \eta_c$           && 4.6   &  $0.8 \pm 0.2 $   \\
$\eta_c{\, '}$   &  $\gamma  {\rm J}/\psi $ && 7.9   &                   \\
\Endtabular
\Endtable

A well-known problem is evident in the
decay rate J$/\psi \to \gamma \eta_c$, which is that the predicted rate
in the nonrelativistic potential model is about a factor of 2-3 larger
than experiment. Since this rate only involves the charm quark magnetic moment, and hence only its mass, this discrepancy is a surprise. The relativized
Godfrey-Isgur model \cite{Godfrey:xj} predicts a somewhat smaller rate
of 2.4~keV, which is still about a factor of two larger than experiment.
Since the errors are rather large, it would clearly be very interesting 
to improve the experimental accuracy of this surprising partial width. 
If this discrepancy is confirmed, it may be an indication that 
pure-$c\bar c$ models are a rather inaccurate description of
charmonium, 
and that other components of the state vector 
such as two-meson continua
make comparable important contributions to the M1 transition amplitudes.
In view of the inaccuracy of the theoretical 
1S M1 transition rate, it would also be interesting to test the 
2S transition rate $\psi{\, '} \to \gamma \eta_c{\, '}$ 
experimentally. Unfortunately, this rate
is predicted to be a rather small 0.21~keV in the
nonrelativistic $c\bar c$ model. 

An even larger discrepancy between experiment and theory is evident in the
``hindered" M1 transition $\psi{\, '} \to \gamma \eta_c$.
Since this rate is only nonzero due to recoil effects (not included here)
and corrections to the naively orthogonal 1S and 2S $c\bar c$
wavefunctions, the discrepancy is perhaps less surprising than that 
found in the allowed 1S $\to$ 1S $\psi \to \gamma \eta_c$ transition rate. 
In any case this is another example of
an M1 transition rate in charmonium in which experiment and theory
are clearly in disagreement. Since the experimental rate is again only
about $4\sigma $ from zero, it would be very useful to improve the accuracy
of this measurement at BES or CLEO-c.

M1 decays between charmonium resonances have only been observed 
between S-wave states.
The rates between orbitally-excited states are typically predicted to be 
quite small, due to the small splittings within excited-L multiplets. 
They are 
large enough however to be observable given narrow initial 
states and large event samples. For example, a hypothetical 
$^1$D$_2$ $c\bar c$ assignment for the X(3872) could be  
tested through a search for its M1 decay to $\gamma \psi(3770)$, 
which has a partial width of 1.2~keV in our
nonrelativistic potential model. The $h_c$ (assumed at 3525~MeV) 
decay $h_c \to \gamma  \chi_0$ has similar phase space, and 
is predicted to have a partial width of 0.8~keV. 
In contrast, the smaller phase space of the M1 transition
from the higher-mass $\chi_2$ state leads to an expected  
partial width for $\chi_2 \to \gamma h_c$ of only about 60~eV. 

\vskip 0.5cm

\subsection{$e^+e^-$ widths of $1^{--}$ states}

Leptonic partial widths are trivially accessible at $e^+e^-$ machines,
and they provide interesting (and currently rather puzzling) information
regarding the wavefunctions of $1^{--}$ charmonium states. 
In the nonrelativistic 
$c\bar c$ potential model $e^+e^-$ annihilation through the photon couples 
to the $c\bar c$ pair at contact, so these widths are only predicted to be 
nonzero for S-wave $c\bar c$ systems. 
This nonrelativistic 
partial width is given by the van Royen - Weisskopf formula 
\cite{VanRoyen:nq},
\be
\Gamma_{c\bar c}^{e^+e^-}({}^3{\rm S}_1) =
\frac{16}{9}\, \alpha^2\; 
\frac{|\psi(0)|^2} {{\rm M}_{c\bar c}^2}     
\ee
where the radial wavefunction is normalized to 
$\int_0^\infty r^2 dr\, |\psi(r)|^2 = 1$. 
Strictly speaking, for relativistic bound states the annihilation
is not completely local, so some coupling of D-wave states to 
$e^+e^-$ is predicted. This width at leading
order is proportional to
$|\psi''(0)|^2$, 
\be
\Gamma_{c\bar c}^{e^+e^-}({}^3{\rm D}_1) =
\frac{50}{9}\, \alpha^2\; 
\frac{|\psi''(0)|^2} {{\rm M}_{c\bar c}^2 m_c^4}     
\ee
which is much smaller than the corresponding widths
of the $^3$S$_1$ states.
Evaluating these widths using the nonrelativistic quark model 
wavefunctions described here gives the results quoted in \Tab{leptonic}. 

\Begtable{ht}
{Predictions of the nonrelativistic $c\bar c$ potential model 
for $e^+e^-$ partial widths, together with current PDG 
experimental values \cite{PDG2004}.} 
{leptonic}
\Begtabular{cccr}
State                    
& Asst.
&   $\Gamma^{e^+e^-}_{thy.}$ (keV): 
& $\Gamma^{e^+e^-}_{expt.}$  (keV):  \\
\hline
J/$\psi$                        
& ${1}\,^3$S$_1$
&  12.13\phantom{1}  
& $ 5.40 \pm 0.17 $    \\
$\psi{\, '}$                         
& ${2}\,^3$S$_1$
&   \phantom{1}5.03\phantom{1}  
& $ 2.12 \pm 0.12 $    \\
$\psi(3770)$                    
& ${1}^3$D$_1$
&   \phantom{1}0.056 
& $ 0.26 \pm 0.04 $    \\
$\psi(4040)$                    
& ${3}\,^3$S$_1$
&   \phantom{1}3.48\phantom{1}  
& $ 0.75 \pm 0.15 $    \\
$\psi(4159)$                    
& ${2}^3$D$_1$
&   \phantom{1}0.096 
& $ 0.77 \pm 0.23 $    \\
$\psi(4415)$                    
& ${4}\,^3$S$_1$
&   \phantom{1}2.63\phantom{1}  
& $ 0.47 \pm 0.10 $    \\
\Endtabular
\Endtable

Note that the agreement between the model and experiment is not 
especially good.
This may be in part due to the choice of parameters, although the 
overestimate of the J/$\psi$ leptonic width by roughly a factor of
two seems to be a common difficulty in naive potential models. 
Non-valence components in the charmonium states, such as $c\bar c g$
or D meson pairs from decay loops, may also contribute to this inaccuracy.
Concerns have also been expressed that the leading-order 
QCD radiative corrections may be large \cite{Barbieri:1975ki}, which could 
significantly reduce the overall scale of the leptonic widths. 
Of course the leading-order corrections are prescription-dependent, 
so it is not clear that this claim is reliable.
In any case it is evident that a simple change of scale 
will not resolve the discrepancies in \Tab{leptonic}, since the 
nominally D-wave states $\psi(3770)$ and $\psi(4159)$ both
have significant leptonic widths. 

The large experimental 
leptonic widths of the $\psi(3770)$ and $\psi(4159)$ relative to predictions
for pure D-wave $c\bar c$ states (see \Tab{leptonic}) may be due
to an admixture of an S-wave $c\bar c$ component
\cite{Rosner:2001nm,Rosner:2004mi}.
As we noted previously, this mixing can be tested 
through measurements of the E1 radiative decay rates 
$\psi(3770) \to \gamma \chi_2$ 
and (with more difficulty)
$\psi(4159) \to \gamma \chi_2$ and $\gamma \chi_2{\, '}$, since these
transitions are very sensitive to the presence of S-wave $c\bar c$ components.
Radiative width ratios such as 
$\Gamma(\psi(3770) \to \gamma \chi_2)/\Gamma(\psi(3770) \to \gamma \chi_1)$ 
and the 
leptonic width ratio 
$\Gamma(\psi(3770) \to e^+e^-)/\Gamma(\psi{\, '} \to e^+e^-)$
provide two independent tests of S-D mixing.

\vskip 0.5cm

\subsection{Two-photon couplings}

Although neither BES nor CLEO-c will have adequate $\sqrt{s}$ to usefully
exploit two-photon production of charmonium resonances, this has been a useful
technique at higher-energy $e^+e^-$ facilities, which we briefly mention here
in the interest of completeness. This subject has been reviewed elsewhere
\cite{Barnes:1992sg}.
                                                                                
Two-photon resonance production involves the reaction
$e^+e^- \to e^+e^- R$, $R\to f$, where $R$ is a C = $(+)$ meson resonance
and $f$ is the exclusive final state observed.
This process proceeds dominantly through the two-photon coupling
of the resonance, $e^+e^- \to e^+e^- \gamma \gamma$, $\gamma \gamma \to R$,
and hence implicitly gives the two-photon partial width
$\Gamma_{\gamma\gamma}(R)$ times the branching fraction
$B(R\to f)$. One may also consider the case of one or both
photons significantly off mass shell $q_\gamma^2 \neq 0$, which gives the
generalized widths $\Gamma_{\gamma^*\gamma}(R)$ and
$\Gamma_{\gamma^*\gamma^*}(R)$. The interesting question for off-shell
widths is whether they are given by a vector dominance formula with the
mass of the relevant vector, in this case the J$/\psi$.
                                                                                
In the limit of large quark mass (and hence a zero-range charm quark
propagator) the two-photon width of an S-wave charmonium state is proportional
to the wavefunction at contact squared. 
In this approximation, higher-L states are
produced with amplitudes proportional to the L$^{th}$ derivative
of the wavefunction at contact \cite{Ackleh:1991ws}.
(In practice this attractively simple contact approximation appears
marginal at best at the $c\bar c$ mass scale.)
Both the $\eta_c$ and $\eta_c{\, '}$ have been seen
in two-photon collisions; the $\gamma\gamma$ width of the
$\eta_c$ is approximately 7~keV, comparable to quark model expectations. The
$\eta_c{\, '}\to \gamma\gamma$ width is not known because one measures the
$\gamma\gamma$ width times the branching fraction to an exclusive final state,
and the $\eta_c{\, '}$ absolute branching fractions are not known.
Higher-L $c\bar c$ two-photon widths are suppressed by the
mass of the charm quark, and therefore are not very well established;
only the P-wave $\chi_0$ and $\chi_2$ states have been observed.
The theoretical ratio of P-wave $\gamma\gamma$ widths
in the large quark mass limit is
$\Gamma_{\gamma\gamma}(^3$P$_0)/
\Gamma_{\gamma\gamma}(^3$P$_2) = 15/4$, however this
may be modified significantly by QCD radiative corrections and finite
quark mass effects. Nonrelativistically the $\gamma\gamma$ state
produced by the $^3$P$_2$ should be pure helicity two;
relativisitic corrections are expected to give rise to a small helicity-zero
$\gamma\gamma$ component as well.
                                                                                
Although one can in principle produce higher-L $c\bar c$ states in
$\gamma\gamma$ collisions, in practice these rates fall rapidly with
increasing L for $c\bar c$ and heavier $Q\bar Q$ systems.
As an example, the two-photon width of a hypothetical
$^1$D$_2$(3840) $c\bar c$ state is predicted to be only 20~eV
\cite{Barnes:1992sg}.
                                                                                
\vskip 0.5cm

\section{Open-charm strong decays}

The dominant decay modes of most hadrons are open-flavor strong decays,
in which the initial quarks separate into different final states. 
Despite their importance in QCD, these open-flavor decays are not at all
well understood on a fundamental level. There are very exciting prospects
for greatly improving our understanding of this important aspect of 
nonperturbative QCD through studies of the open-flavor decays of charmonium
states above the DD threshold. 

Strong decay couplings are also of great interest 
because at second-order they give rise to 
$(c\bar c) \to (c\bar q)(q\bar c) \to (c\bar c)$ 
loop diagrams, which are not included in pure $c\bar c$ potential models
or in quenched lattice gauge theory. As these virtual decay loop effects 
are believed to be numerically important in charmonium
\cite{vanBeveren:bd,Heikkila:1983wd}, 
it is correspondingly important to improve our 
understanding of the strong decay couplings that drive these processes. 
This will be most straightforward through studies of the decay couplings of
$c\bar c$ states above open-charm threshold.   

At present these open-flavor decays are treated
using simple quark model descriptions of the decay process, which 
are known to give numerically reasonable results for many light hadron 
decays. The earliest of these decay models is the 
``$\, {}^3{\rm P}_0$ model" of Micu~\cite{Micu:1968mk}
and 
LeYaouanc~{\it et al.}~\cite{LeY73}, 
who describe the decay in terms of the local
production of a new $q\bar q$ with $0^{++}$ quantum numbers 
(hence ${}^3{\rm P}_0$). 
The overall strength of this dimensionless pair production amplitude 
$\gamma$ is simply fitted to experiment, and for light mesons is found
to be $\gamma \approx 0.4$ \cite{Barnes:1996ff}.

The best known theoretical work on charmonium is the series of papers
by the Cornell group 
\cite{Eichten:1974af,Eichten:ag,Eichten:1978tg,Eichten:1979ms},
which includes a study of
open-flavor strong decays \cite{Eichten:1979ms}.
Remarkably, the strong decay model assumed by the Cornell group is
{\it not} the standard ${}^3{\rm P}_0$ model. Instead, they assume that strong
decays take place through pair production from the linear confining potential,
which they assume to transform as the time component 
of a Lorentz vector, $\gamma_0$. 
(Vector confinement was assumed in some hadron models in the 1980s, but has 
since been largely replaced by the assumption of Lorentz scalar confinement.)

A more recent study of possible QCD strong decay mechanisms by 
Ackleh~{\it et al.} 
\cite{Ackleh:1996yt}
found that the ${}^3{\rm P}_0$ model decay 
amplitudes were quite similar in form to what one would find for 
pair production by the linear scalar confining interaction, although the
usual string tension leads to an overestimated decay amplitude by about
a factor of 2-3 for light meson decays. 

These two models of the decay interaction, Lorentz vector versus Lorentz
scalar, will give different predictions for open-flavor strong decay 
amplitudes that can be tested at BES and CLEO-c. 
Unfortunately the Cornell group
did not give detailed amplitude decompositions for their Lorentz vector
decay model; instead they evaluated a ``resolvent" that 
gave the cross section ratio $R$, and showed graphical results for $R$
and the branching fractions. The required evaluation of decay amplitudes 
in this model is in progress \cite{Swa04}. Similarly, 
the corresponding ${}^3{\rm P}_0$ decay model amplitudes have (remarkably) 
never been evaluated in detail for the higher charmonium states. This work 
has now largely been completed \cite{Bar04}, and some of the results are 
presented here. In particular we will discuss results specific to the 
easily accessible $1^{--}$ states.

The results we find for the open-charm decay modes of the four 
accessible $1^{--}$ states in the ${}^3{\rm P}_0$ model, with
SHO wavefunctions and parameters of 
$\beta = 0.5$~GeV (length scale) and $\gamma=0.4$ 
(pair production amplitude) are shown in \Tab{OCdecays_table}. 
The lightest
state is the $\psi(3770)$, which is usually 
considered a ${}^3{\rm D}_1$ state, can only decay to
DD; it is predicted to have a total width
of 43~MeV, about twice the experimental value.
The total widths of the nominally 
3${}^3{\rm S}_1$ $\psi(4040)$ 
and
2${}^3{\rm D}_1$ $\psi(4159)$ 
are both predicted to be rather near the PDG experimental 
values \cite{PDG2004};
74~MeV (theory) versus $52\pm 10$~MeV (experiment) for the $\psi(4040)$,
and
73~MeV (theory) versus $78\pm 20$~MeV (experiment) for the $\psi(4159)$.  

The branching fractions of the $\psi(4040)$ are a famous problem; although
phase space favors DD over DD$^*$, and D$^*$D$^*$ has essentially no
phase space, the existing data actually suggests that the relative 
branching fractions are D$^*$D$^* >> $DD$^*  >> $ DD. 
(The ratios quoted by Mark~I at SLAC were
$32.0\pm 12.0 : 1 : 0.05\pm 0.03$ 
\cite{PDG2004,Goldhaber:1977qn}.)
The ${}^3{\rm P}_0$ model
is at least partially in agreement here, as is the Cornell decay model 
\cite{Eichten:1979ms}, 
since
both predict a very small DD mode. (In the decay models this is a 
result of a node in the decay amplitude of the radially-excited 
3${}^3{\rm S}_1$ state.) 
The Cornell decay model however predict comparable 
branching fractions to  DD$^*$ and D$^*$D$^*$. The ${}^3{\rm P}_0$ model
with our parameters finds a similar result (see \Tab{OCdecays_table}),
although there is a node in the DD$^*$ decay amplitude that can also
significantly reduce the DD$^*$ branching fraction \cite{Bar04}. 
It would be very interesting to measure and recalculate 
these branching fractions carefully to determine whether there 
is indeed disagreement with theoretical expectations. Note also that
the D$^*$D$^*$ mode is especially interesting, in that there are three
amplitudes in this final state, 
${}^1{\rm P}_1$, ${}^5{\rm P}_1$ and ${}^5{\rm F}_1$.
The ${}^3{\rm P}_0$ model predicts
that decays from S-wave $c\bar c$ states 
have zero ${}^5{\rm F}_1$ amplitudes, and that the ratio of the 
two P-wave amplitudes is ${}^5{\rm P}_1/{}^1{\rm P}_1 = -2\sqrt{5}$. 

The 2${}^3{\rm D}_1$ candidate $\psi(4159)$ has the same modes 
accessible as the $\psi(4040)$, but is predicted to have a very different 
pattern of branching fractions and decay amplitudes. The mode DD$^*$ 
is theoretically strongly suppressed, 
DD should be large, and  D$^*$D$^*$ should be the leading mode. 
The relative amplitudes within the 
D$^*$D$^*$ final state are predicted to have a very characteristic 
pattern; ${}^5{\rm F}_1$ is predicted to be dominant, and 
${}^1{\rm P}_1$ is predicted to be larger than ${}^5{\rm P}_1$, in the 
ratio ${}^1{\rm P}_1 / {}^5{\rm P}_1 = -\sqrt{5}$.

Finally, the $\psi(4415)$ is assigned to 4${}^3{\rm S}_1$ in potential models. 
This highly
radially-excited level has several decay amplitude nodes that make the 
predicted decay branching fractions less reliable. The expected 
dominant mode among S-wave pairs is D$^*$D$^*$, with the same amplitude
ratios quoted above for the 3${}^3{\rm S}_1$ state.

An interesting
new feature of $\psi(4415)$ decays is that there is 
sufficient phase space for decays to S+P D-meson pairs; 
in particular, the interesting new CLEO states
D$^*_{\rm sJ}(2317)$ and
D$_{\rm s1}(2460)$ can both be made in S-wave 
(assuming that the ${\rm D}_{\rm sJ}(2317)$) is a scalar), in
$\psi(4415) \to {\rm D}^*_s {\rm D}^*_{\rm sJ}(2317)$ and
$\psi(4415) \to {\rm D}_s {\rm D}_{\rm s1}(2460)$ respectively.
(Nominally these decay modes are just above threshold, but they are
accessible once the width of the $\psi(4415)$ is incorporated.)
Since the branching fractions expected to these final states are 
not especially small \cite{Bar04}, the $\psi(4415)$ could provide a 
straightforward source of a large sample of 
D$^*_{\rm sJ}(2317)$ and D$_{\rm s1}(2460)$ events.

\section{Conclusions}

In this note we have reviewed some topics of current interest 
in charmonium 
which can be addressed at the BES and CLEO-c facilities. 
The topics discussed here were electromagnetic couplings and decays 
(E1, M1, $e^+e^-$ and two-photon), hadron decay loops 
and open-flavor strong decays. 
Studies of properties of the $\psi(3770)$,
$\psi(4040)$ and $\psi(4159)$ are especially interesting in this regard.
The E1 decays of the $\psi(3770)$, in particular to $\gamma \chi_2$,
can be compared with the $e^+e^-$ width to allow tests of S-D mixing,
which may be due to open-flavor decay loops. The relative strong
branching fractions of the $\psi(4040)$ and the $\psi(4159)$ to
DD, DD$^*$ and D$^*$D$^*$,
and in particular the relative size of the three
D$^*$D$^*$ decay amplitudes, allow sensitive tests of strong decay models.
Finally, the high-mass tail of the $\psi(4415)$ may be useful as a source
of events containing the new narrow resonances
D$^*_{\rm sJ}(2317)$ and D$_{\rm s1}(2460)$.

\section*{Acknowledgments}

I would like to thank the organizers of the Beijing BES/CLEO-c joint workshop,
in particular Wei-Guo Li and Ian Shipsey,
for their kind invitation to review aspects of charmonium at this meeting,
and for the opportunity to discuss physics with my fellow participants.
This report
is an extended version of material presented at the workshop.
I am also grateful to F.E.Close, S.Godfrey, J.Quigg, J.Rosner and
E.S.Swanson for discussions of various aspects of the material presented here.

\vfill\eject

\Begtable{ht}
{Predictions of the nonrelativistic $c\bar c$ potential model 
discussed in the text for the charmonium spectrum to 4.2~GeV 
(and for all S-wave states to 4.5~GeV).} 
{spec_table}
\Begtabular{cccc}
State                    &   Mass (MeV): & Expt. (input)  & Theor.  \\
\hline
J/$\psi$                    &&  3097 & 3090    \\
$\eta_c$                    &&  2979 & 2982    \\
\hline
$\psi{\, '}$                &&  3686 & 3672     \\
$\eta_c{\, '}$                   &&  3638 & 3630    \\
\hline
$\psi(3^3{\rm S}_1)$        &&  4040 & 4072     \\
$\eta_c(3^1{\rm S}_0)$      &&       & 4043    \\
\hline
$\psi(4^3{\rm S}_1)$        &&  4415 & 4406     \\
$\eta_c(4^1{\rm S}_0)$      &&       & 4384    \\
\hline
$\chi_2$                    &&  3556 & 3556    \\
$\chi_1$                    &&  3511 & 3505    \\
$\chi_0$                    &&  3415 & 3424    \\
$h_c$                       &&       & 3516    \\
\hline
$\chi_2(2^3{\rm P}_2)$      &&       & 3972    \\
$\chi_1(2^3{\rm P}_1)$      &&       & 3925    \\
$\chi_0(2^3{\rm P}_0)$      &&       & 3852    \\
$h_c  (2^1{\rm P}_1)$       &&       & 3934    \\
\hline
$\psi_3   ({}^3{\rm D}_3)$  &&       & 3806    \\
$\psi_2   ({}^3{\rm D}_2)$  &&       & 3800    \\
$\psi     ({}^3{\rm D}_1)$  && 3770  & 3785    \\
$\eta_{c2}({}^1{\rm D}_2)$  &&       & 3799    \\
\hline
$\psi_3   (2^3{\rm D}_3)$   &&       & 4167    \\
$\psi_2   (2^3{\rm D}_2)$   &&       & 4158    \\
$\psi     (2^3{\rm D}_1)$   && 4159  & 4142    \\
$\eta_{c2}(2^1{\rm D}_2)$   &&       & 4158    \\
\hline
$\chi_4   ({}^3{\rm F}_4)$  &&      & 4021    \\
$\chi_3   ({}^3{\rm F}_3)$  &&      & 4029    \\
$\chi_2   ({}^3{\rm F}_2)$  &&      & 4029    \\
$\eta_{c3}({}^1{\rm F}_3)$  &&      & 4026    \\
\Endtabular
\Endtable

\vfill\eject

\Begtable{ht}
{Theoretical and experimental E1 radiative partial widths 
of the easily accessible $1^{--}$ 
states, as well as some interesting additional cases (see text). 
The experimental $\psi(3770)$ numbers are taken from 
Ref.\cite{Rosner:2004mi}.
}
{rad_table}
\Begtabular{ccccc}
Initial state & Final state   
& \phantom{xxxx}  &  $\Gamma_{thy.}$~(keV) & $\Gamma_{expt.}$~(keV) \\ 
\hline
$\chi_2$   & $\gamma {\rm J}/\psi $   && 424  &  $426 \pm 48$  \\
$\chi_1$   &  $\gamma {\rm J}/\psi$   && 320  &  $288 \pm 51$  \\
$\chi_0$   &  $\gamma {\rm J}/\psi$   && 155  &  $119 \pm 17$  \\
$h_c$      &  $\gamma\eta_c$          && 494  &                \\
\hline
$\psi{\, '}$    &  $\gamma\chi_2$    && 38  &  $18.0 \pm 2.0$  \\
           &  $\gamma\chi_1$    && 54  &  $23.6 \pm 2.7$  \\
           &  $\gamma\chi_0$    && 62  &  $24.2 \pm 2.5$  \\
$\eta_c{\, '}$  &  $\gamma h_c$      && 49  &  \\
\hline
$\psi(3^3{\rm S}_1)(4040) $  &  $\gamma\chi_2$    && 0.5  &    \\
$                         $  &  $\gamma\chi_1$    && 0.4  &    \\
$                         $  &  $\gamma\chi_0$    && 0.2  &    \\
$                         $  &  $\gamma\chi_2(2^3{\rm P}_2)$    && 14  &    \\
$                         $  &  $\gamma\chi_1(2^3{\rm P}_1)$    && 39  &    \\
$                         $  &  $\gamma\chi_0(2^3{\rm P}_0)$    && 54  &    \\
\hline
$\psi({}^3{\rm D}_1)(3770)$  &  $\gamma\chi_2$    && 4.9  
& $\leq 330$ (90\% c.l.)  \\
$                         $  &  $\gamma\chi_1$    && 126  
& $280 \pm 100$           \\
$                         $  &  $\gamma\chi_0$    && 405  
& $320 \pm 100$            \\
\hline
$\psi(2^3{\rm D}_1)(4159) $  &  $\gamma\chi_2$    &&  0.8  &    \\
$                         $  &  $\gamma \chi_1$   && 14    &    \\
$                         $  &  $\gamma \chi_0$   && 27    &    \\
$                         $  &  $\gamma\chi_2(2^3{\rm P}_2)$    &&   5.9  &    \\
$                         $  &  $\gamma\chi_1(2^3{\rm P}_1)$    && 168    &    \\
$                         $  &  $\gamma\chi_0(2^3{\rm P}_0)$    && 485    &    \\
$                         $  &  $\gamma\chi_2({}^3{\rm F}_2)$   &&  51    &    \\
\Endtabular
\Endtable

\vfill\eject

\Begtable{ht}
{Open-charm strong decay modes of the $1^{--}$ states accessible at 
BES and CLEO-c. A reaction-dependent factor has been removed from the 
decay subamplitudes in the final column of the table, so only the amplitude 
ratios are physically meaningful.}
{OCdecays_table}
\Begtabular{llccl}
State & Mode & 
$\Gamma_{expt.}$ (MeV) 
& $\Gamma_{thy.}$ (MeV) 
& Subamps. 
\\ \hline
$\psi(3770)$ 
(${}^3$D$_1$)
&  DD 
&  
& 43. 
& 
\\   
&  all  
& $23.6 \pm 2.7 $ 
& 43. 
&      
\\
\hline
$\psi(4040)$   
($3\, {}^3$S$_1$)
&  DD 
& 
& 0.1
& 
\\
&  DD$^*$ 
& 
& 33. 
& 
\\
&  D$_s$D$_s$ 
& 
& 8.
& 
\\
&  D$^*$D$^*$ 
& 
& 33. 
& ${}^1{\rm P}_1 = +0.056  $   
\\
&   
& 
&  
& ${}^5{\rm P}_1 = -0.251  $   
\\
&   
& 
&  
& ${}^5{\rm F}_1 = 0  $   
\\
&  all  
& $52 \pm 10 $ 
& 74. 
&      
\\
\hline
$\psi(4159)$   
($2\, {}^3$D$_1$)
&  DD 
& 
& 16.
& 
\\
&  DD$^*$ 
& 
& 0.4
& 
\\
&  D$^*$D$^*$ 
& 
& 35. 
& ${}^1{\rm P}_1 = +0.081  $   
\\
&   
& 
&  
& ${}^5{\rm P}_1 = -0.036  $   
\\
&   
& 
&  
& ${}^5{\rm F}_1 = -0.141  $   
\\
&  D$_s$D$_s$ 
& 
& 8.
& 
\\
&  all  
& $78 \pm 20 $ 
& 73. 
&      
\\
\hline
$\psi(4415)$   
($4\, {}^3$S$_1$)
&  DD 
& 
& 0.4
& 
\\
&  DD$^*$ 
& 
& 2.3 
& 
\\
&  D$^*$D$^*$ 
& 
& 16. 
& ${}^1{\rm P}_1 = -0.018  $   
\\
&   
& 
&  
& ${}^5{\rm P}_1 = +0.081  $   
\\
&   
& 
&  
& ${}^5{\rm F}_1 = 0  $   
\\
&  D$_s$D$_s$ 
& 
& 1.3
& 
\\
&  D$_s$D$_s$$^*$ 
& 
& 2.6 
& 
\\
&  D$_s$$^*$D$_s$$^*$ 
& 
& 0.7 
& ${}^1{\rm P}_1 = +0.006  $   
\\
&   
& 
&  
& ${}^5{\rm P}_1 = -0.028  $   
\\
&   
& 
&  
& ${}^5{\rm F}_1 = 0  $   
\\
&  S+P modes~\cite{Bar04}  
& 
& 
&      
\\
&  all  
& $52 \pm 10 $ 
& 
&      
\\
\Endtabular
\Endtable

\vfill\eject


\begin{thebibliography}{99}


\bibitem{Choi:2002na}
S.~K.~Choi {\it et al.}  [BELLE collaboration],
Phys.\ Rev.\ Lett.\  {\bf 89}, 102001 (2002)
[Erratum-ibid.\  {\bf 89}, 129901 (2002)]
[arXiv:hep-ex/0206002].

\bibitem{Aubert:2003pt}
B.~Aubert {\it et al.}  [BABAR Collaboration],
arXiv:hep-ex/0311038.

\bibitem{unknown:2003bk}
J.~Ernst  {\it et al.}  [CLEO Collaboration],
arXiv:hep-ex/0306060.
                                                                                
\bibitem{Liao:2002rj}
X.~Liao and T.~Manke,
arXiv:hep-lat/0210030.
         
\bibitem{vanBeveren:bd}
E.~van Beveren, C.~Dullemond and G.~Rupp,
Phys.\ Rev.\ D {\bf 21}, 772 (1980)
[Erratum-ibid.\ D {\bf 22}, 787 (1980)].

\bibitem{Heikkila:1983wd}
K.~Heikkila, S.~Ono and N.~A.~Tornqvist,
Phys.\ Rev.\ D {\bf 29}, 110 (1984)
[Erratum-ibid.\ D {\bf 29}, 2136 (1984)].

\bibitem{Kwong:1988ae}
W.~Kwong and J.~L.~Rosner,
Phys.\ Rev.\ D {\bf 38}, 279 (1988).
                                                                                
\bibitem{Godfrey:xj}
S.~Godfrey and N.~Isgur,
Phys.\ Rev.\ D {\bf 32}, 189 (1985).
                                                                                
\bibitem{Rosner:2001nm}
J.~L.~Rosner,
Phys.\ Rev.\ D {\bf 64}, 094002 (2001)
[arXiv:hep-ph/0105327].
                                                                                
\bibitem{Rosner:2004mi}
J.~L.~Rosner,
arXiv:hep-ph/0405196.
                                                                                
\bibitem{Bar04}
T.Barnes, S.Godfrey and E.S.Swanson (in preparation).
                                                                                
\bibitem{VanRoyen:nq}
R.~Van Royen and V.~F.~Weisskopf,
Nuovo Cim.\ A {\bf 50}, 617 (1967)
[Erratum-ibid.\ A {\bf 51}, 583 (1967)].

\bibitem{PDG2004}
S.Eidelman \etal (Particle Data Group),
Phys. Lett. B592, 1 (2004).
                                                                                
\bibitem{Barbieri:1975ki}
R.~Barbieri, R.~Gatto, R.~Kogerler and Z.~Kunszt,
Phys.\ Lett.\ B {\bf 57}, 455 (1975).
                                                                                
\bibitem{Barnes:1992sg}
T.~Barnes,
``Two photon couplings of quarkonia with arbitrary J(PC),''
{\it Int. Workshop on Photon-Photon Collisions,
La Jolla, CA, Mar 22-26, 1992.}\\
http://www.slac.stanford.edu/spires/find/hep/www?r=ornl-ccip-92-05
                                                                                
\bibitem{Ackleh:1991ws}
E.~S.~Ackleh, T.~Barnes and F.~E.~Close,
Phys.\ Rev.\ D {\bf 46}, 2257 (1992).
                                                                                
\bibitem{Micu:1968mk}
L.~Micu,
Nucl.\ Phys.\ B {\bf 10}, 521 (1969).
                                                                                
\bibitem{LeY73}
A.LeYaouanc, L.Oliver, O.P\`ene and J.Raynal,
Phys. Rev. D8, 2223 (1973);
{\it ibid.}, D9, 1415 (1974); D11, 680 (1975); D11, 1272 (1975);
Phys. Lett. 71B, 397 (1977).
                                                                                
\bibitem{Barnes:1996ff}
T.~Barnes, F.~E.~Close, P.~R.~Page and E.~S.~Swanson,
Phys.\ Rev.\ D {\bf 55}, 4157 (1997)
[arXiv:hep-ph/9609339].
                                                                                
\bibitem{Eichten:1974af}
E.~Eichten, K.~Gottfried, T.~Kinoshita, J.~B.~Kogut, K.~D.~Lane and T.~M.~Yan,
Phys.\ Rev.\ Lett.\  {\bf 34}, 369 (1975)
[Erratum-ibid.\  {\bf 36}, 1276 (1976)].
                                                                                
\bibitem{Eichten:ag}
E.~Eichten, K.~Gottfried, T.~Kinoshita, K.~D.~Lane and T.~M.~Yan,
Phys.\ Rev.\ Lett.\  {\bf 36}, 500 (1976).
                                                                                
\bibitem{Eichten:1978tg}
E.~Eichten, K.~Gottfried, T.~Kinoshita, K.~D.~Lane and T.~M.~Yan,
Phys.\ Rev.\ D {\bf 17}, 3090 (1978)
[Erratum-ibid.\ D {\bf 21}, 313 (1980)].
                                                                                
\bibitem{Eichten:1979ms}
E.~Eichten, K.~Gottfried, T.~Kinoshita, K.~D.~Lane and T.~M.~Yan,
Phys.\ Rev.\ D {\bf 21}, 203 (1980).
                                                                                
\bibitem{Ackleh:1996yt}
E.~S.~Ackleh, T.~Barnes and E.~S.~Swanson,
Phys.\ Rev.\ D {\bf 54}, 6811 (1996)
[arXiv:hep-ph/9604355].
                                                                                
\bibitem{Swa04}
E.S.Swanson (in preparation).
                                                                                
\bibitem{Goldhaber:1977qn}
G.~Goldhaber {\it et al.},
Phys.\ Lett.\ B {\bf 69}, 503 (1977).
                                                                                
\end{thebibliography}
\end{document}